# Resilience of the Spin-Orbit Torque against Geometrical Backscattering

Seng Ghee Tan[(1) †], Che — Chun Huang[(2)], Mansoor B. A. Jalil[(3)], Ching — Ray Chang[(1)],

Szu — Cheng Cheng[(1)]

(1) Department of Optoelectric Physics, Chinese Culture University, 55 Hwa-Kang Road, Yang-Ming-Shan, Taipei 11114, Taiwan

(2) Department of Physics, National Taiwan University, Taipei 10617, Taiwan

(3) Department of Electrical and Computer Engineering, National University of Singapore, 4 Engineering Drive 3, Singapore 117576

Abstracts

We show in this paper that the technologically relevant field-like spin-orbit torque shows resilience against the geometrical effect of electron backscattering. As device grows smaller in sizes, the effect of geometry on physical properties like spin torque, and hence switching current could place a physical limit on the continued shrinkage of such device - a necessary trend of all memory devices (MRAM). The geometrical effect of curves has been shown to impact quantum transport and topological transition of Dirac and topological systems. In our work, we have ruled out the potential threat of line-curves degrading the effectiveness of spin-orbit torque switching. In other words, spin-orbit torque switching will be resilient against the influence of curves that line the circumferences of defects in the events of electron backscattering, which commonly happen in the channel of modern electronic devices.

Contacts:
Seng Ghee Tan (Prof)
Department of Optoelectric Physics,
Chinese Culture University,
55, Hwa-Kang Road, Yang-Ming-Shan,
Taipei, Taiwan 11114 ROC
(Tel: 886-02-2861-0511, DID:25221)

* Email: csy16@ulive.pccu.edu.tw ; tansengghee@gmail.com

PACS:

**Introduction**

Recently, the physics of curves has been studied for its effect on many modern quantum systems, e.g. the Dirac-electronic transport due to the topological surface states **[1-4]**. Curves were also studied for their effects on topological transitions, from inducing topological phases **[5]** in a Rashba system to shifting band-inversion point **[6]** in the BHZ topological-insulator system. In spintronics **[7,8]**, the effect of curve on spin and charge transport has been studied in spin-orbit systems like the Rashba and the Dresselhaus. More specifically, the effect of surface curves was studied for the vertical spin polarization that it generates **[9]** - the same spin effect responsible for spin Hall **[10-14]** and the damping-like spin-orbit torque (SOT) **[15]**. As the technology of spin-transfer torque (STT) magnetic memory (MRAM) matures, sights were set on the SOT as a future switching mechanism. This paper is dedicated to studying the physics of curve due to electron tracing the circumferences of defects on a specific type of SOT known as the field-like SOT **[16-21]**, first discussed in 2007 **[16]** and experimentally confirmed later **[20, 21].** We show that the technologically relevant field-like SOT is resilient against these geometrical effects in the course of electron backscattering.

As device grows smaller in sizes, the effect of geometry on physical properties like spin torque, and hence switching current could place a physical limit on the continued shrinkage of device sizes, which is a necessary trend of the memory industry that seeks to pack ever more device elements into a standard IC footprint. Indeed, as mentioned above, curves have been shown to have significant impact on topological transport and transitions, clearly affecting their Berry curvatures due to band-structure. It remains to be seen if they might pose significant physical impact on spin torque, which also has origins in Berry curvature and spin-orbit coupling. While the physics of curve can impact device physics locally, we emphasize again that our focus in this work is on the global effect (physical property summed spatially over the entire device), which has real device relevance with respect to total current. In other words, the current required to switch the magnetization in a SOT-MRAM device is therefore theoretically stable against curves of defects which might be abundantly present in devices.

When a structure is described as curved here, it means that electron traveling in this structure is constrained in the dimension of its motion. For illustration, we consider a concept 2D structure based on a standard SOT-MRAM, in which a curve (a round arc) is inscribed on the surface such that electron motion is constrained to one specific dimension, e.g. to trace the external circumference of the arc in a clockwise fashion. The actual SOT-MRAM is a multilayer hetero-structure with a ferromagnetic (FM) layer deposited on top of a heavy atom (HA) layer like the Pt or Pd. The Rashba spin-orbit coupling is assumed at the FM-HA surface. The FM layer can be Fe, Co, and so forth, and hosts a magnetization vector that points along $\theta^M$ with the $x$ axis.

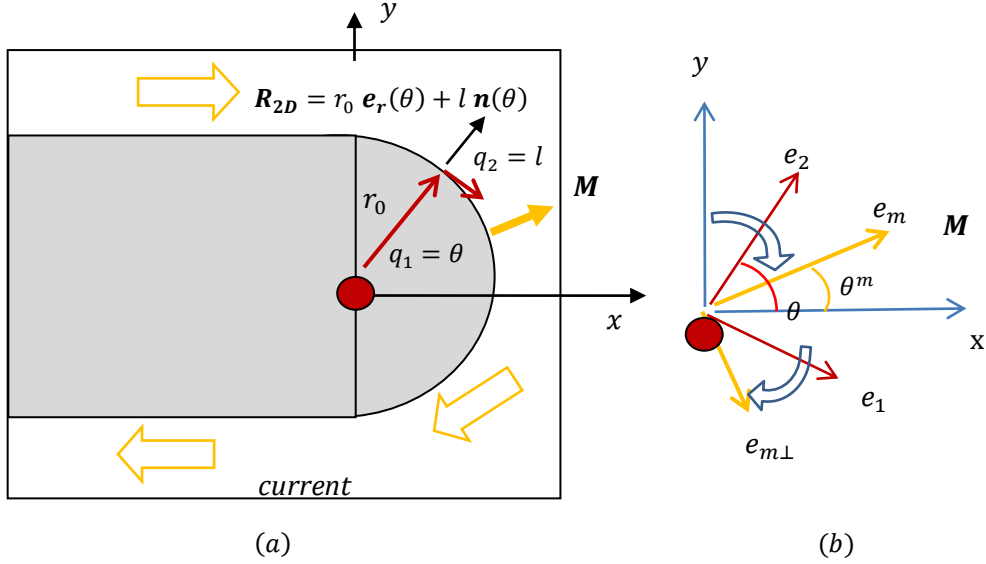

FIG.1. (a) Drone's view of a 2D structue with an inscribed in-plane circular arc (to mimic defects) such that current traces the circumference of the arc externally. (b) Directions of magnetization vector with respect to the Cartesian coordinates and the reparameterized coordinates.

Electron transport which traces the circumference of the concept structure above mimics backscattering due to defects commonly present in the central region (channel) of 2D devices as shown below. Figure 2(a) shows an electron tracing the circumference in a clockwise manner, while Fig.2(b) shows an anti-clockwise tracing. Electrons scattering off of a defect (Fig.2(c) & Fig.2(d)) are considered to have not traced any significant part of the defect curve. For simplicity, defects are represented by smooth circular structures and electrons are assumed to track the macroscopic roundness but not the microscopic jaggedness.

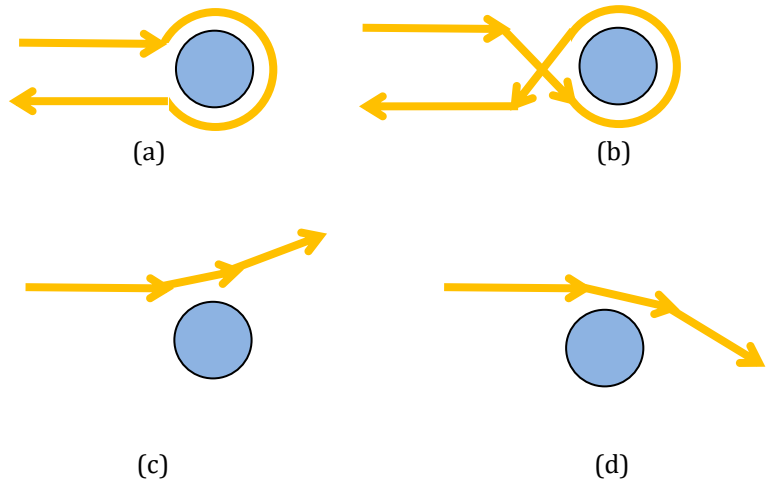

Fig.2. (a) Backscattering of an electron going around the defect in a clockwise manner; (b) Backscattering of an electron going around the defect in an anti-clockwise manner. (c) & (d) Electron scatters off of a defect.

The SOT-MRAM with defects can be viewed as a system that comprises the kinetic, spin-orbit coupling, and magnetic energies along a line-curve. The full Hamitonian in Hermitian form that describes the device is given by

$$H = \frac{1}{2m} P_a P^a + \frac{1}{2m}(P_a S^a + S^a P_a) + \sigma^a m_a$$

(1)

The term $(P_a S^a + S^a P_a)$ is Hermitian when $P$ and $S$ are individually Hermitian. A minimal coupling form is presented below to more intuitively reflect the "forceful" effect on the electron, arising due to spin-orbit coupling

$$H = \frac{1}{2m}(P_a + S_a)(P^a + S^a) + g_{ab}\sigma^a m^b$$

(2)

where $\frac{1}{2m}P_a P^a = -\frac{i\hbar}{2m}(\partial_a + \partial_a \ln\sqrt{g})P^a$. This is because when $P_a$ is quantized, the operator will be a co-variant derivative with $\Gamma_a = \partial_a \ln\sqrt{g}$ playing the role of the Christoffel symbol in classical curve physics, and the Hamiltonian will be

$$H = -\frac{i\hbar}{2m}(\partial_a + \Gamma_a)P^a \psi - \frac{i\hbar}{2m}(\partial_a + \Gamma_a)S^a \psi + \frac{1}{2m}S^a P_a + g_{ab}\sigma^a m^b$$

(3)

The explicit expression of *g* would depend on the exact geometry of the curve and the choice of coordinates. In the following, we have chosen the device surface to contain the *x-y* axis with the *z* axis being vertical to the surface. The *x, y* coordinates are generalized to $e_1, e_2$ which are contained in the same *x-y* plane. An arbitrary in-plane spatial point can therefore be accessed anywhere with vector equation $R_{2D} = r(q_1) + q_2 n(q_1)$ (origin marked in red) as shown in Fig.1. Vector $n(q_1)$ would be perpendicular to the tangent of the arbitrary in-plane curve at any point. Specifically, the vector equation for a round curve of $x^2 + y^2 = r_0^2$ would be $R_{2D} = r_0 e_r(\theta) + l\, n(\theta)$. For a curve of $x^2 + cy = r_0^2$, the vector equation is $R_{2D} = r_0 e_r(\theta) + \left(\frac{r_0^2}{c}\sin^2\theta - r_0 \sin\theta\right)e_y + l\,n(\theta)$. In all cases, the coordinate sets are: $(q_1 = \theta, q_2 = l)$ and $(e_1 = e_\theta, e_2 = e_l)$, and the vector equations retain form $R_{2D} = r(q_1) + q_2 n(q_1)$. For simplicity, $e_\theta$ is known as the surfaced direction, $e_l$ the radial. The vector equations are required to find the metric tensor *g* which quantifies the curvature effect on the physical properties of those curves, e.g. conductance, permittivity, or in our case the SOT. The actual effect of the curve would therefore be reflected in *g*. For example, to represent defects with round structues as shown in Fig.2., *g* can be determined as follows

$$g_{ij} = \begin{bmatrix} g_{\theta\theta} & 0 \\ 0 & 1 \end{bmatrix} = \begin{bmatrix} (r_0 + l)^2 & 0 \\ 0 & 1 \end{bmatrix} \quad ; \quad g^{ij} = \begin{bmatrix} \frac{1}{g} & 0 \\ 0 & 1 \end{bmatrix}$$

(4)

where $\sqrt{g} = (\partial_\theta^y \partial_l^x - \partial_\theta^x \partial_l^y)$, and $\partial_b^a$ stands for $\frac{\partial a}{\partial b}$. To study the magnetic and spin-orbit physics on line-curves, one performs a local gauge transformation in the spin space such that the local frame would be rotated to "some" axis – the choice of which pretty much determines the physics to be revealed from those energies.

$$H' = -\frac{i\hbar}{2m}\left(\partial_a + \Gamma_a + \frac{ie}{\hbar}A_a\right)(P^a + eA^a) - \frac{i\hbar}{2m}\left(\partial_a + \Gamma_a + \frac{ie}{\hbar}A_a\right)US^aU^\dagger$$
$$- US^aU^\dagger \frac{i\hbar}{2m}\left(\partial_a + \frac{ie}{\hbar}A_a\right) + g_{ab}U\sigma^a m^b U^\dagger$$

(5)

where $A_a = -i\frac{\hbar}{e}U(\partial_a U^\dagger)$. For simplicity, one can define $\nabla_a \equiv \partial_a + \Gamma_a$. Note that the transformed Hamiltonian is still general in the sense that we haven't decided which axis the local frame should be rotated to. It is therefore clear that the physics that one would like to "see" lies in this important decision, i.e. the choice of the transformation operator $U$. By experience, SOT physics would become apparent in a transformation that rotates the $y$ axis to the magnetization axis $\boldsymbol{e_m}$, and $x$ axis to $\boldsymbol{e_{m\perp}}$, all in the presence of spin-orbit coupling, as shown in Fig.1b above. Mathematically the above is satisfied by the choice of $U$ that performs

$$U\boldsymbol{\sigma}U^\dagger = \sigma^y \boldsymbol{e_m} + \sigma^x \boldsymbol{e_{m\perp}},$$

(6)

which leads to

$$U\boldsymbol{\sigma}.\boldsymbol{m}\, U^\dagger = \sigma^y.$$

(7)

Upon completing the transformation, local gauge potentials would appear in the Hamiltonian. Rewriting Eq. (5) in a more symmetrical form, one has

$$H' = \left(\frac{1}{2m}\right)(-i\hbar\nabla_a + eA_a + US_aU^\dagger)(-i\hbar\partial^a + eA_a + US^aU^\dagger) + g_{ab}U\sigma^a m^b U^\dagger$$

(8)

The physics of curve, spin-orbit coupling, and magnetism is now captured in the expression of a total gauge: $\mathbb{V}_a = eA_a + US_aU^\dagger$. In fact, the rotation gauge $eA_a$ has been associated with a form of adiabatic spin tansfer torque, which was studied previously **[22]** and would not be further elaborated in this article. As our interest is in the SOT, the relevant term will be the transformed spin-orbit gauge **[13, 16, 19, 23]**, also known henceforth as $\mathcal{F}_a = US_a U^\dagger$. The relevant energies are:

$$E = \psi^\dagger \left(\frac{1}{2m}\right)(-i\hbar\nabla_a + \mathcal{F}_a)(-i\hbar\partial^a + \mathcal{F}^a)\psi$$

(9)

Dropping the kinetic energy terms, the scope is further narrowed to our focus on just the effect of SOT in the backscattering process

$$E_{int} = \left(\frac{-i\hbar}{2m}\right)\psi^\dagger[\nabla_a \mathcal{F}^a + \mathcal{F}_a \partial^a]\psi$$

(10)

Recalling the term $\frac{1}{2m}(P_a S^a + S^a P_a)$ in Eq.(1), it is evident that similar Math stucture is preserved in $(\nabla_a \mathcal{F}^a + \mathcal{F}_a \partial^a)$ which ensures the Hermiticity of the overall expression as long as the individual operators of Eq.(10) are Hermitian. Now, we will examine each energy density $(\nabla_a \mathcal{F}^a, \mathcal{F}_a \partial^a)$ term in details:

$$\nabla_a \mathcal{F}^a \to \left(\frac{-i\hbar}{2m}\right)[\partial_a(\psi^\dagger \mathcal{F}^a \psi) - (\partial_a \psi^\dagger)\mathcal{F}^a \psi + \Gamma_a (\psi^\dagger \mathcal{F}^a \psi)]$$

(11)

$$\mathcal{F}_a \partial^a \to \left(\frac{-i\hbar}{2m}\right)\psi^\dagger \mathcal{F}_a \partial^a \psi$$

(12)

The expression $\nabla_a \mathcal{F}^a$ is further reduced to $\left(\frac{-i\hbar}{2m}\right)[-(\partial_a\psi^\dagger)\mathcal{F}^a\psi]$ as the first and the third term on the RHS vanish as surface terms when the energy density is integrated over the entire device as follows

$$\sqrt{g}\int[\partial_a + \Gamma_a](\psi^\dagger \mathcal{F}^a \psi)\, dV = \sqrt{g}\int(\psi^\dagger \mathcal{F}^a \psi).dS = 0$$

(13)

The above simply means that explicit curve physics in the well-known form of Christoffel symbol $(\Gamma_a = \partial_a ln\sqrt{g})$ vanishes over the entire device. The Christoffel effect could, however, be physically signifiant in the local sense. That would depend on further treatment that should also look into the Hermiticity of these individual terms. But as far as the total current for spin torque and magnetization is concerned, local curve effect is not important if it vanishes globally. It therefore suffices to proceed with the remaining terms which combine to make physical sense as follows,

$$\left(\frac{-i\hbar}{2m}\right)[\mathcal{F}_a\psi^\dagger\partial^a\psi - (\partial_a\psi^\dagger)\mathcal{F}^a\psi] \leftrightarrow j_a\mathcal{F}^a$$

(14)

where $j_a$ has the physical meaning of current density. It is worth noting that the expression above remarkably preserves once again the Math structure of $\frac{1}{2m}(P_a S^a + S^a P_a)$. Since $\mathcal{F}^a$ orginates from the spin-orbit coupling in a curved space, the entire term $j_a \mathcal{F}^a$ would be an interaction energy density related to the coupling of the current flow ($j_a$) with the curved spin-orbit physics. In the following, we will adopt the curvilinear coordinates $(\theta, l)$. Here $j_a$ comes from the presumption of the plane-wave curvilinear eigenstate for a curved mesoscopic structure, in the same way $\psi = e^{iky}$ is generally assumed as the transverse plane-wave solution for a flat mesoscopic structure. To evaluate this physial term in a more intuitive manner, and recalling that $\mathcal{F}_a = U S_a U^\dagger$, the spin-orbit gauge in curvilinear coordinates will be

$$\mathcal{F}_\theta = -U\left(\frac{\alpha\, m}{\hbar}\sqrt{g}\sigma^l\right)U^\dagger, \quad \mathcal{F}_l = U\left(\frac{\alpha\, m}{\hbar}\sqrt{g}\sigma^\theta\right)U^\dagger$$

(15)

where explicitly $S^\theta = -\frac{\alpha\, m}{\hbar\sqrt{g}}\sigma_l$, $S^l = \frac{\alpha\, m}{\hbar\sqrt{g}}\sigma_\theta$ and $S_\theta = -\frac{\alpha\, m}{\hbar}\sqrt{g}\sigma^l$, $S_l = \frac{\alpha\, m}{\hbar}\sqrt{g}\sigma^\theta$. Note also that $S_\theta = g_{\theta\theta}S^\theta + g_{\theta l}S^l$. We take a digression to review the derivation of the explicit $\boldsymbol{S_a}$. Here we begin with

$$H_{soc} = \frac{\alpha}{\hbar}\boldsymbol{\sigma}\cdot(\boldsymbol{P}\times\boldsymbol{e_3}) = \frac{\alpha}{\hbar}\sigma^v \boldsymbol{e_v}\cdot\left[P^k \boldsymbol{e_k}\times\left(\frac{\boldsymbol{e_1}\times\boldsymbol{e_2}}{\sqrt{g}}\right)\right]$$

(16)

which leads to

$$H_{soc} = -\frac{\alpha}{\hbar}\sqrt{g}\,\sigma^2 P^1 + \frac{\alpha}{\hbar}\sqrt{g}\,\sigma^1 P^2 = \frac{P^1}{m}S_1 + \frac{P^2}{m}S_2$$

(17)

Refering to Eq.(1) where $H_{soc} = \frac{1}{2m}(P_a S^a + S^a P_a)$, the above is merely its first part, i.e. $P_a S^a$. As already elaborated, the second part $S^a P_a$ is only required to satisfy Hermiticity. For the studies of spin-orbit coupling, it suffices to consider just the first part. It can then be deduced that $\boldsymbol{S_a} = \frac{\alpha\, m}{\hbar\sqrt{g}}\boldsymbol{e_3}\times\boldsymbol{\sigma^b}$, with the definition of $\boldsymbol{\sigma^1} = \sigma^1 \boldsymbol{e_1}$. Note also that $\boldsymbol{n} = \boldsymbol{e_1}\times\boldsymbol{e_2} = \sqrt{g}\,\boldsymbol{e_3}$. One would now have the energy density

$$E_{int} = j_a \mathcal{F}^a = \left(\frac{\alpha m}{e\hbar}\sqrt{g}\right)\left[-j^\theta\, U\sigma^l U^\dagger + j^l\, U\sigma^\theta U^\dagger\right]$$

(18)

As long as constrain has not been imposed, the choice of basis is only for illustration as they can be converted to one another straightforwardly. For example, in the Cartesian coordinates, the above would simply be

$$E_{int} = j_a \mathcal{F}^a = \left(\frac{\alpha m}{e\hbar}\right)[-j^y\, U\sigma^x U^\dagger + j^x\, U\sigma^y U^\dagger]$$

(19)

with $\sqrt{g} \to 1$. The advantage of the curvilinear coordinate is in the convenience of considering a curve effect or imposing constrain. For example, with curve effect the radial current is set to zero ($j^l \to 0$), and the constrained energy would simply be

$$E_{int} = \left(\frac{\alpha m}{e\hbar}\sqrt{g}\right)[-j^\theta\, U\sigma^l U^\dagger]$$

(20)

This would be equivalent to setting the follwing: $q_2 = 0$, as well as $\Gamma_1(q_1, q_2 = 0)$ and $\Gamma_2(q_1, q_2 = 0)$. But as stated earlier, $\Gamma_a$ had vanished before the constrain and are therefore inconsequential. Back to the free system, we consider a potential in the magnetic space to arise from the energy density $E_{int}$. The potential is essentially the effective anisotropy field as it is commonly understood in micro-magnetic physics

$$\mu\, \boldsymbol{H} = \frac{\delta E_{int}}{\delta \boldsymbol{M}} = \frac{\delta}{\delta \boldsymbol{M}}\left(\frac{j^a \mathcal{F}_a}{e}\right)$$

(21)

The explicit expression of $\mu\, \boldsymbol{H}$ is therefore given as

$$\mu\, \boldsymbol{H} = \left(\frac{\alpha\, m}{\hbar}\sqrt{g}\right)\left[\frac{-j^\theta}{e}\left(\frac{\delta}{\delta \boldsymbol{M}} U\sigma^l U^\dagger\right) + \frac{j^l}{e}\left(\frac{\delta}{\delta \boldsymbol{M}} U\sigma^\theta U^\dagger\right)\right]$$

(22)

We will now look into the spin physics and focus on the Pauli matrices. Originally expressed in curvilinear coordinates $(\theta, l)$, the Pauli matrices are re-expressed in the Cartesian coordinates $(x, y)$ in which the adiabatic physics will be introduced. Conversion of physical quantities into different basis is provided in Table 1. Note that $\sqrt{g} = \partial_l^x \partial_\theta^y - \partial_l^y \partial_\theta^x$ can be derived from $g = (g_{\theta\theta}g_{ll} - g_{\theta l}g_{l\theta})$.

Table 1. This table shows the dimension conversion of a physical quantity from $(\theta, l)$ coordinates to Cartesian coordinates.

|  | Dimension 2: Radial | Dimension 1: Surface |
|---|---|---|
| Magnetic moment | $\sqrt{g}m^l = \partial_\theta^y m^x - \partial_\theta^x m^y$ <br> $m_l = \dfrac{\partial_y^\theta m_x - \partial_x^\theta m_y}{\partial_x^l \partial_y^\theta - \partial_y^l \partial_x^\theta}$ | $\sqrt{g}m^\theta = \partial_l^x m^y - \partial_l^y m^x$ <br> $m_\theta = \dfrac{\partial_x^l m_y - \partial_y^l m_x}{\partial_x^l \partial_y^\theta - \partial_y^l \partial_x^\theta}$ |
| Pauli Matrix | $\sqrt{g}\sigma^l = \partial_\theta^y \sigma^x - \partial_\theta^x \sigma^y$ | $\sqrt{g}\sigma^\theta = \partial_l^x \sigma^y - \partial_l^y \sigma^x$ |
| Current Density | $\sqrt{g}j^l = \partial_\theta^y j^x - \partial_\theta^x j^y$ | $\sqrt{g}j^\theta = \partial_l^x j^y - \partial_l^y j^x$ |

The effective anisotropy field is now

$$\mu\, \boldsymbol{\delta H} = \left(\frac{\alpha m}{\hbar M_s}\sqrt{g}\right)\left[-\frac{j^\theta}{e}\frac{\delta}{\delta \boldsymbol{n}}\left(\frac{\partial_\theta^y U\sigma^x U^\dagger - \partial_\theta^x U\sigma^y U^\dagger}{\sqrt{g}}\right) + \frac{j^l}{e}\frac{\delta}{\delta \boldsymbol{n}}\left(\frac{\partial_l^x U\sigma^y U^\dagger - \partial_l^y U\sigma^x U^\dagger}{\sqrt{g}}\right)\right]$$

(23)

Note that in the context of a standard SOT device for the technology of magneic memory, $\alpha$ is the Rashba spin-orbit coupling constant. The vacuum permeability is given by $\mu = 4\pi \times 10^{-7}\frac{T}{Am^{-1}}$ or $NA^{-2}$. It is then clear that $\mu\, \boldsymbol{\delta H}$ is like a magnetic field with unit of Tesla. The magnetization is given by $M = \frac{moment}{volume} = \frac{Am^2}{m^3} = Am^{-1}$, which shares the same unit as $\boldsymbol{\delta H}$. On the RHS, $[j][A][M]^{-1} = Tesla$. We will now take a digression to examine the dimension of the spin-orbit gauge and the current terms. The surface current density term $\left(\frac{j^\theta}{e}\right)$ has the dimension of $\left[\frac{nv}{l}\right]$, while gauge term $(U\sigma^l U^\dagger)$ is dimensionless. Likewise, radial current density $\left(\frac{j^l}{e}\right)$ has the dimension of $[nv]$, while gauge $(U\sigma^\theta U^\dagger)$ has the dimension of $\left[\frac{1}{l}\right]$. On the other hand, $(\sqrt{g})$ has the dimension of $[l]$, resulting in $\left(\frac{\alpha m}{\hbar}\sqrt{g}\right)$ having the dimension of $[p\, l]$. The entire RHS would thus have the dimension of $\left[n\frac{E}{M}\right]$, noting also that $n$ has the dimension of $\left[\frac{1}{l^3}\right]$. To find the explicit expressions of $U\sigma^x U^\dagger$, and $U\sigma^y U^\dagger$, one recalls the transformation that was imposed at the outset, i.e. $U\boldsymbol{\sigma}U^\dagger = \sigma^y \boldsymbol{e_m} + \sigma^x \boldsymbol{e_{m\perp}}$. We will now write the magnetization unit vectors $\boldsymbol{e_m}$ and $\boldsymbol{e_{m\perp}}$ in terms of Cartesian coordinates

$$\boldsymbol{e_m} = \boldsymbol{e_x}\cos\theta^m + \boldsymbol{e_y}\sin\theta^m = n^x \boldsymbol{e_x} + n^y \boldsymbol{e_y}$$

$$\boldsymbol{e_{m\perp}} = \frac{\partial \boldsymbol{e_m}}{\partial \theta^m} = -\boldsymbol{e_x}\sin\theta^m + \boldsymbol{e_y}\cos\theta^m = -n^y \boldsymbol{e_x} + n^x \boldsymbol{e_y}$$

(24)

Refering to $U\boldsymbol{\sigma}U^\dagger = \sigma^y \boldsymbol{e}_m + \sigma^x \boldsymbol{e}_{m\perp}$, one can now proceed as follows

$$U\boldsymbol{\sigma}U^\dagger = \sigma^y(\boldsymbol{e}_x \cos\theta^m + \boldsymbol{e}_y \sin\theta^m) + \sigma^x(-\boldsymbol{e}_x \sin\theta^m + \boldsymbol{e}_y \cos\theta^m)$$
$$= (\sigma^y n^x - \sigma^x n^y)\boldsymbol{e}_x + (\sigma^y n^y + \sigma^x n^x)\boldsymbol{e}_y \qquad (25)$$

In fact the above is just a restatement of $U\boldsymbol{\sigma}U^\dagger = U\sigma^x U^\dagger \boldsymbol{e}_x + U\sigma^y U^\dagger \boldsymbol{e}_y$. The results are substituted into Eq.(23) to show the effective field with the Pauli matrices in *x-y* basis

$$\mu\,\boldsymbol{H} = \left(\frac{\alpha m}{\hbar M_s}\sqrt{g}\right)\left[\left(\frac{-j^\theta}{e}\right)\left(\frac{\partial_\theta^y(\sigma^y \boldsymbol{e}_x - \sigma^x \boldsymbol{e}_y)}{\sqrt{g}} - \frac{\partial_\theta^x(\sigma^x \boldsymbol{e}_x + \sigma^y \boldsymbol{e}_y)}{\sqrt{g}}\right)\right.$$
$$\left. + \left(\frac{j^l}{e}\right)\left(\frac{\partial_l^x(\sigma^x \boldsymbol{e}_x + \sigma^y \boldsymbol{e}_y)}{\sqrt{g}} - \partial_l^y\frac{(\sigma^y \boldsymbol{e}_x - \sigma^x \boldsymbol{e}_y)}{\sqrt{g}}\right)\right] \qquad (26)$$

At this point, an important physical step is taken, i.e. the adiabatic approximation in which it is assumed that electron spin aligns along the new *y* axis that has been rotated parallel to $\boldsymbol{e}_m$. Under such spin alignment whereby $\langle y|\sigma^x|y\rangle = 0$, $\langle y|\sigma^y|y\rangle = 1$,

$$\mu\,\boldsymbol{H}_F = \left(\frac{\alpha m}{\hbar M_s}\sqrt{g}\right)\left[\left(\frac{-j^\theta}{e}\right)\boldsymbol{a}_l + \left(\frac{j^l}{e}\right)\boldsymbol{a}_\theta\right] \qquad (27)$$

where $\boldsymbol{a}_l = \left(\frac{\partial_\theta^y \boldsymbol{e}_x - \partial_\theta^x \boldsymbol{e}_y}{\sqrt{g}}\right)$, and $\boldsymbol{a}_\theta = \left(\frac{\partial_l^x \boldsymbol{e}_y - \partial_l^y \boldsymbol{e}_x}{\sqrt{g}}\right)$. In summary, we have performed a local gauge transformation in curvilinear coordinates and derived the effective field in that basis. The Pauli matrices were re-expressed in *x-y* basis and were eliminated through the adiabatic physics, resulting in Eq.(27) which is now a classical effective magnetic field denoted with $\mu\,\boldsymbol{H}_F$. This classical field encapsulates the physics of SOT and plays the role of magnetization switching. Just prior to constraining the electron motion, we perform a quick check to see that the same effective field can be retrieved in *x-y*. We referred once again to Table 1, and obtained

$$\mu\,\boldsymbol{H}_F = \left(\frac{\alpha m}{\hbar M_s}\sqrt{g}\right)\left[\left(\frac{-j^\theta}{e}\right)\boldsymbol{a}_l + \left(\frac{j^l}{e}\right)\boldsymbol{a}_\theta\right] = \left(\frac{\alpha m}{e\hbar M_s}\right)[-j^y \boldsymbol{e}_x + j^x \boldsymbol{e}_y], \qquad (28)$$

which is identical to the well-known field-like SOT effective magnetic field **[16-21]**. Imposing the constrain would require one to set the radial current to zero as $j^l \to 0$, which gives $\sqrt{g}j^l = \partial_\theta^y j^x - \partial_\theta^x j^y = 0$. This is a useful relation that can be substituted into the constrained effective field $\mu\,\boldsymbol{H}_{FC} = \left(\frac{\alpha m}{\hbar M_s}\sqrt{g}\right)\left[\left(\frac{-j^\theta}{e}\right)\boldsymbol{a}_l\right]$ to lead once again to

$$\mu\,\boldsymbol{H}_{FC} = \left(\frac{\alpha m}{e\hbar M_s}\right)[-j^y \boldsymbol{e}_x + j^x \boldsymbol{e}_y].$$

(29)

We have reached an interesting result, i.e. $\mu\,\boldsymbol{H}_F = \mu\,\boldsymbol{H}_{FC}$. At first glance, the above simply shows that eliminating the radial current $\left(\frac{j^l}{e}\right)\boldsymbol{a}_\theta$ "by force" from the equation has no effect on the physics of the effective field. But in addition, the fact that $\mu\,\boldsymbol{H}_F = \mu\,\boldsymbol{H}_{FC}$ also means that $\left(\frac{j^l}{e}\right)\boldsymbol{a}_\theta$ is naturally absent, and therefore eliminating it via motion constrain is an inconsequential process. To explicitly demonstrate this, careful use is made of $\sqrt{g}j^l = \partial_\theta^y j^x - \partial_\theta^x j^y$ (without setting it to zero) in the square brackets of Eq.(30) below, which shows the natural cancellation of the $j^l$ terms,

$$\mu\,\boldsymbol{H}_F = \left(\frac{\alpha m}{\hbar M_s e}\right)\left(\frac{1}{\sqrt{g}}(-\partial_l^x j^y\,\partial_\theta^y \boldsymbol{e}_x + [\partial_l^y j^y\,\partial_\theta^x \boldsymbol{e}_x + \partial_l^y \sqrt{g}j^l \boldsymbol{e}_x]) - j^l(\partial_l^y)\boldsymbol{e}_x\right)$$

$$+ \left(\frac{\alpha m}{\hbar M_s e}\right)\left(\frac{1}{\sqrt{g}}(-\partial_l^y j^x\,\partial_\theta^x \boldsymbol{e}_y + [\partial_l^x j^x\,\partial_\theta^y \boldsymbol{e}_y - \partial_l^x \sqrt{g}j^l \boldsymbol{e}_y]) + j^l(\partial_l^x)\boldsymbol{e}_y\right)$$

(30)

The above leads finally to

$$\mu\,\boldsymbol{H}_F = \left(\frac{\alpha m}{\hbar M_s e}\frac{1}{\sqrt{g}}\right)(-\partial_l^x\,\partial_\theta^y + \partial_l^y\,\partial_\theta^x)j^y \boldsymbol{e}_x + \left(\frac{\alpha m}{\hbar M_s e}\frac{1}{\sqrt{g}}\right)(\partial_l^x\,\partial_\theta^y - \partial_l^y\,\partial_\theta^x)j^x \boldsymbol{e}_y$$
$$= \left(\frac{\alpha m}{e\hbar M_s}\right)[-j^y \boldsymbol{e}_x + j^x \boldsymbol{e}_y]$$

(31)

which is what is indeed $\mu\,\boldsymbol{H}_{FC} = \left(\frac{\alpha m}{\hbar M_s}\sqrt{g}\right)\left[\left(\frac{-j^\theta}{e}\right)\boldsymbol{a}_l\right]$. Once again as far as the physics of effective field of the SOT is concerned, geometrical resilience against electron backscattering around defects is demonstrated.

**Conclusion**
We set out to study the effect of curves on the physics of SOT, an important physical quantity for the technology of future memory. It was found that when integrated over the entire device, SOT shows a resilience against the influence of geometry in the course of electron backscattering. The crucial physics comes from the disappearance of the momentum-coupled Christoffel symbols as surface terms upon a global summation. However, as the transformed spin-orbit gauge, which captures the physics of spin-orbit coupling, magnetic moment, and geometrical curve, is still in the picture, it remains unclear if geometry would still have influence on SOT. A careful analysis which involves studying the adiabatic physics in the lab coordinates finally shows that the effective field in curvilinear coordinates remains unaffected by the elimination of the radial current – a manual process to reflect constrain on the system. In fact, the effective field is solely a function of the surface current. Forceful removal of the radial current is therefore inconsequential. SOT remains resilience against curves that line the circumferences of defects on device surfaces.

**Acknowledgement**
We would like to thank the Ministry of Science and Technology of Taiwan for supporting this work under Grant. No.: 107-2112-M-034-002-MY3.

**References**
[1] Yositake Takane, Ken-Ichiro Imura, J. Phys. Soc. Japan **82**, 074712 (2013).
[2] Ken-Ichiro Imura, Yositake Takane, and Akihiro Tanaka, Phys. Rev. B **84**, 195406 (2011).
[3] V. Parente, P. Lucignano, P. Vitale, A. Tagliacozzo, and F. Guinea, Phys. Rev. B **83**, 075424 (2011).
[4] Dung-Hai Lee, Phys. Rev. Letts. **103**, 196804 (2009).
[5] Paola Gentile, Mario Cuoco, Carmine, Phys. Rev. Letts. **115**, 256801 (2015).
[6] Zhuo Bin Siu, Jian-Yuan Chang, Seng Ghee Tan, Mansoor B. A. Jalil, Ching-Ray Chang, Sci. Reps. **8**, 16497 (2018).
[7] Tai-Chung Cheng, Jiann-Yeu Chen, Ching-Ray Chang, Phys. Rev. B **84**, 214423 (2011).
[8] Ming-Hao Liu, Jhih-Sheng Wu, Son-Hsien Chen, and Ching-Ray Chang, Phys. Rev. B **84**, 085307 (2011).
[9] Zhuo Bin Siu, Mansoor B. A. Jalil, and Seng Ghee Tan, J. Appl. Phys. **121**, 233902 (2017).
[10] S. Murakami, N. Nagaosa, S.C. Zhang, Science **301**, 1348 (2003).
[11] T. Fujita, M. B. A. Jalil, S. G. Tan, J. Phys. Soc. Jpn. **78**, 104714 (2009).
[12] T. Fujita, M. B. A. Jalil, S. G. Tan, New J. Phys. **12**, 013016 (2010).
[13] T. Fujita, M. B. A. Jalil, S. G. Tan, Shuichi Murakami, J. Appl. Phys. [Appl. Phys. Rev.] **110**, 121301 (2011).
[14] Congson Ho, Seng Ghee Tan, Mansoor B. A. Jalil, EPL **107**, 37005 (2014).
[15] H. Kurebayashi et al., Nature Nanotechnology **9**, 211 (2014).
[16] S. G. Tan, M. B. A. Jalil, and Xiong-Jun Liu, arXiv:0705.3502, (2007).
[17] K. Obata and G. Tatara, Phys. Rev. B **77**, 214429 (2008).
[18] A. Manchon, S. Zhang, Phys. Rev. B **78**, 212405 (2008).
[19] S. G. Tan, M. B. A. Jalil, T. Fujita, and X. J. Liu, Ann. Phys. (NY) **326**, 207 (2011).
[20] Ioan Mihai Miron et al., Nat. Mater. **9**, 230 (2010).
[21] Jun Yeon Kim et al., Nature Materials **12**, 240 (2013).
[22] Ya. B. Bazaliy, B. A. Jones, Shou-Cheng Zhang, Phys. Rev. B **57**, R3213 (1998).
[23] S. G. Tan, M. B. A. Jalil, Xiong-Jun Liu, and T. Fujita, Phys. Rev. B **78**, 245321 (2008).